# *Can thermal input from a prior universe account for relic graviton production?*

A.W. Beckwith
*abeckwith@UH.edu*

## Abstract

The author presents how answering Sean Carroll's supposition of a pre inflation state of low temperature-low entropy pre inflation state as given in 2005 provides a bridge between two models with radically different predictions. I.e. Loop quantum gravity may be giving us a template as to thermal input which answers in the affirmative if or not relic graviton production can exist. Brane world models as constructed by Randall and Sundrum permit the low entropy conditions Sean Carroll and Jennifer Chen predicted in 2005, and then of course how to go from the brane world model so outlined to the 10 to the 32 Kelvin conditions stated by Weinberg in 1972 as necessary for quantum gravity . Afterwards, we would have a transition to Guth style inflation. This is also a way of getting around the get around the fact that conventional cosmological CMB is limited by a barrier as of a red shift limit of about z = 1000, i.e. when the universe was about 1000 times smaller and 100,000 times younger than today as to photons. It also allows a working model of quintessence scalar fields of a very short term time scale on the order of Planck time magnitude which allows for relic graviton production and leads to dark matter/dark energy values predicted via current Chapyron joint dark matter/dark energy models after Guth style chaotic inflation.

# I. INTRODUCTION

First of all we need to consider if there is an inherent fluctuation in early universe cosmology which is linked to a vacuum state nucleating out of 'nothing'. The answer we have is yes and no. The vacuum fluctuation leads to production of a dark energy density which we can state is initially due to contributions from an axion wall, which is dissolved during the inflationary era. What we will be doing is to reconcile how that wall was dissolved in early universe cosmology with quantum gravity models, brane world models, and Weinberg's published as of 1972 prediction of a threshold of 10 to the 32 power Kelvin for quantum effects becoming dominant in quantum gravity models. All of this is leading up to conditions in which we can expect relic graviton production which could account for the presence of strong gravitational fields in the onset of Guth style inflation, which would be in line with Penrose's predictions via the Jeans inequality as to low temperature, low entropy conditions for pre inflationary cosmology.

Still though, as written up by L. Crowell in 'Quantum Fluctuations in Space-Time'[18], we can make a first principle argument as to the uncertainty principle in quantum mechanics being derivable from variations in the space time metric, and especially in the $g_{00}$ component of the metric $g_{ij}$, i.e. $L \cdot \delta g \approx \delta g_{00} \cdot c^2$, where $L$ is a unit length for evaluation of quantum fluctuations. From there we can then argue that fluctuations in dark energy density were initially of the form $\delta \rho_{DE} \approx \sqrt{\delta \rho_\Lambda} \cdot \sqrt{\rho_{Planck\ length}}$ with $\delta \rho_\Lambda$ connected with energy density fluctuations due to variations w.r.t. temperature in the Planck 'constant' parameter in a formulation first presented by Park et al. We are assuming that this variation of the Planck 'constant' parameter is also connected with relic graviton production. And we use a short time interval quintessence equation of 'motion' in order to show how there could be a bridge between a pre inflationary quantum bounce thermal input of energy to initially cooled down states of matter condition postulated by Carroll and J. Chen [15], 2005 to Guth style quadratic potential inflation.

Also, relatively recently , quintessence models were for large time scales ruled out , and T. Padmanabhan, in a 2005 world press scientific tome edited by Abbay Ashtekar, on pp 175-201[51] notes that this is due to the cosmological constant being set to zero for the quintessence models being considered. This is due in part to the classes of potentials offered along the lines of Eqn. 18 of T. Padmanabhan's article [51], which we did not use, as well as the duration of the quintessence period being far larger than Planck's time interval. We got about this by using a different scalar model than what Padmanabhan mentioned and also use a vanishingly small time interval at the onset of the regime just before inflation to have quintessence scalar field evolution, with a collapse of axion walls leading to the chaotic inflationary potential first written up by Guth in the 1980s. As it is, our present paper is in response to suggestions by Dr. Wald [60] (2005), Sean Carroll, and Jennifer Chen [15] (2005) , and others in the physics department in the U. of Chicago about a Jeans instability criteria leading to low entropy states of the universe at the onset of conditions before inflationary physics initiated expansion of inflaton fields. We agree with their conclusions and think it ties in nicely with the argument so presented as to a burst of relic gravitons being produced. This also is consistent with an answer as to the supposition for the formation of a unique class of initial vacuum states, answering a question Guth [28] raised in 2003 about if or not a preferred form of vacuum state for early universe nucleation was obtainable. This is in tandem with the addition of gravity changing typical criteria for astrophysical applications of the jeans instability criteria [58] for weakly interacting fields, as mentioned by Penrose [52].

## II STATEMENT OF THE GENERAL PROBLEM WE ARE INVESTIGATING

Contemporary graviton theory states as a given that there is a thermal upsurge which initiates the growth of graviton physics. This is shown in K.E. Kunzes well written (2002) article [39] which gives an extremely lucid introduction as to early universe additional dimensions giving a decisive impetus to giving additional momentum to the production of relic gravitons. However, Kunze [39] is relying upon enhanced thermal excitation states, which contradict the Jeans instability criteria which appears to rule out a gravitational field soaked initial universe configuration being thermally excited. Is there a way to get around this situation which appears to violate the Jeans instability criteria for gravitational fields/gravitons in the early universe mandating low entropy states? We believe that there is, and that it relies upon a suggestion given by Ashtekar, A., Pawlowski, T. and Singh, P [5,6] (2006) as to the influence of the quantum bounce via quantum loop gravity mirror imaging a prior universe collapsing into a 'singularity' with much the same geometry as the present universe. If this is the case, then we suggest that an energy flux from that prior universe collapse is transferred into low entropy thermally cooled down initial state, leading to a sudden burst of relic gravitons as to our present universe configuration. The first order estimate for this graviton burst comes from the numerical density equation for gravitons written up by Weinberg as of 1971 [61] with an exponential factor containing a frequency value divided by a thermal value, T, minus 1. If the frequency value is initially quite high, and the input given by a prior universe 'bounce', with an initial very high value of energy configuration, then we reason that this would be enough to introduce a massive energy excitation into a thermally cooled down axion wall configuration which would then lead to the extreme temperatures of approximately $10^{12}$ Kelvin forming at or before a Planck interval of time $t_P$, plus a melt down of the axion domain wall, which we then says presages formations of a Guth style inflationary quadratic and the onset of chaotic inflationary expansion.

A way of getting to all of this is to work with a variant of the Holographic principle, and an upper bound to entropy calculations. R.Busso, and L. Randall [14] (2001) give a brane world variant of the more standard upper bounds for entropy in terms of area calculations times powers of either the fourth or fifth dimensional values of Planck mass (45), which still lead to minimized values if we go near the origins of the big bang itself. Our observations are then not only consistent with the upper bound shrinking due to smaller and smaller volume/area values of regions of space containing entropy measured quantities, but consistent with entropy/area being less than or equal to a constant times absolute temperatures, if we take as a given in the beginning low temperature conditions prior to the pop up of an inflation scalar field.

Recently, Bo Feng et. al. [23], introduced the idea of an effective Lagrangian to compliment the idea of CPT violations, as new physics, composed of a term proportional to the derivative of a scalar field $\phi$ (in this case a quintessence field) times the dual of the electromagnetic tensor. What we are supplying is a proof if you will of time dependence of the quintessence scalar field, and Bo Feng et. al. [22] inputs into the electric and magnetic fields of this dual of the E&M tensor from the stand point of CMB. It is noteworthy to bring up that Ichiki et al. [34] notes that because standard electromagnetic fields are coupled to gravity, magnetic fields simply dilute away as the universe expands, i.e. we need to consider the role of gravity generation in early universe models. We will, in this document try to address how, via graviton production, we have intense gravity wave generation, and also how to use this as a probe of early universe quintessence fields, and also how to get around the fact that conventional cosmological CMB is limited by a barrier as of a red shift limit of about $z \approx 1000$, i.e. when the universe was about 1000 times smaller and 100,000 times younger than today as to photons ,i.e. we are confirming as was stated by Weinberg [61] as of 1977 that there is zero chance of relic photon generation from the big bang itself we can see being observable is zero and that we are using relic gravitons as a probe as to the physics of quintessence fields, as well as the origins of dark matter/dark energy issue.

In addition this approach accounts for data suggesting that the four-dimensional version of the "cosmological constant" in fact varies with respect to external background temperature. If this temperature significantly varied during early universe baryogenesis, the end result is that there would be a huge release of spin-two gravitons in the early stages of cosmic nucleation of a new universe. It also answers whether "Even if there are $10^{1000}$ vacuum states produced by String theory, then does inflation produces overwhelmingly one preferred type of vacuum states over the other possible types of vacuum states?" [28] (Guth, 2003).

Also, we also account for the evolution of an equation of motion of a quintessence field, via equations given to us by M. Li, X. Wang, B. Feng, and X. Zhang [43], which is a first ever re do in dept of the interaction of a quintessence scalar field with baryonic 'normal matter' assuming varying contributions to a potential field system with a varying by temperature axion mass contribution to an evolving pre inflationary state, which collapses to a quadratic Guth style inflationary state with a suitable rise in initial inflationary temperatures.

Finally, we can state unequivocally that the typical dark energy SUGRA potential, i.e. for times $0 < t \approx t_P$ or larger

$$V(\phi) = \frac{\Lambda^{\alpha+4}}{\phi^\alpha} \cdot \exp\left(+ 4 \cdot \pi \cdot \phi^2 / m_{Planck}^2\right) \tag{1}$$

blends / over laps well with the finalized Guth style inflationary quadratic potential so derived in our formulation of this problem after we have a thermal input from a quantum bounce which melts axion walls. This being the case only because the intervals of time are so short as to merely indicate that Eqn. 1 merely indicates the growth behavior of a quantum vacuum state. In addition, we should also mention that the super nova survey, called Essence, has experimentally measured the 'dark energy' - the thing that is causing the acceleration of the universe - to better than within 10% accuracy. The feature of this dark energy that was measured is its 'equation of state'. And our work leads to the standard cosmological constant value well within the 1000 year limit of observational accuracy after the big bang , due to thermal cooling off which would occur about the z = 1000 or so red shift limit . This being the case, we will construct the following two tables for outlining the principles involved

Begin with assuming that the absolute value of the five dimensional cosmological 'constant' parameter is inversely related to temperature, i.e.

$$|\Lambda_{5-\dim}| \propto c_1 \cdot (1/T) \tag{2}$$

As opposed to working with the more traditional four dimensional version of the same, minus the minus sign of the brane world theory version

$$\Lambda_{4-\dim} \propto c_2 \cdot T \tag{3}$$

Then we can write, for small time values $t \approx \delta^1 \cdot t_P$, $0 < \delta^1 \leq 1$ and for temperatures sharply lower than $T \approx 10^{12} \, Kelvin$

$$\left(\frac{\Lambda}{|\Lambda_5|} - 1\right) \approx O\left(\frac{1}{n}\right) \sim \textbf{To the order of } (1/n) \tag{4}$$

We can do this for length( radii ) values proportional to the value of the inverse of what the Hubble parameter is when the absolute value of five dimensional cosmological 'constant' parameter is of the order of the four dimensional cosmological 'constant' parameter, i.e. when the critical initial nucleation length we consider obeys

$$L \propto L(\Lambda_{4-\dim} \approx |\Lambda_{5-\dim}|) \propto H^{-1}(\Lambda_{4-\dim} \approx |\Lambda_{5-\dim}|) \tag{5}$$

So being the case, we get a simple case of where we can analyze vacuum energy density fluctuations in the region of space smaller than the radii given in Eqn. (5) above, via

$$\Delta\rho_{vacuum} \propto l_P^{-2} \cdot L^{-2}(\Lambda_{4-\dim} \approx |\Lambda_{5-\dim}|) \propto H^2(\Lambda_{4-\dim} \approx |\Lambda_{5-\dim}|)/G \tag{6}$$

This is with respect to working with dimensions of the order of Planck's length, i.e. the volume of space where the volume of space has a radii which is of the order of, for $0 \leq \widetilde{\alpha} \leq N_+$

$$L(\Lambda_{4-\dim} \approx |\Lambda_{5-\dim}|) \approx 10^{\widetilde{\alpha}} \times (l_P = .1616 \times 10^{-33} \, cm) \tag{7}$$

Where initially we have temperatures of the order of 1.4 times 10 to the 32 power Kelvin as a thresh hold for the existence of quantum effects. This would pre suppose answering the issue raised by Weinberg [61]. As of 1972, he wrote that for quantum effects to be dominant in cosmology, with a value of critical energy we will use in setting a template for relic graviton production later on.

$$E_{critical} \equiv 1.22 \times 10^{28} \, eV \tag{7a}$$

This is pre supposing that we have a working cosmology which actually gets to such temperatures at the instance of quantum nucleation of a new universe. **Appendix Ia.** as accessed below gives us a working format as to the dynamics of quantum nucleation as outlined in this article. **Appendix 1b** which is in the same page gives commentary as to temperature dependence of the cosmological constant. Furthermore, **Appendix II** gives temperature dependence of four and five dimensional versions of a 'cosmogical constant. And if there is no temperature dependence, in the 5$^{th}$ dimensional cosmological constant se set as having magnitude $\Lambda$, we still can get a five dimensional line element [63]

$$dS^2_{5-\dim} \cong \frac{\Lambda \cdot l^2}{3} \cdot \{4-\dim \, Schwartzshield \, deSitter \, metric\} - dl^2 \tag{7b}$$

Finally, we should mention **Appendix III** which describes how a short lived quintessence phenomena leads to initial conditions for a relic graviton surge which does not get into the physical problems long time quintessence creates.

## III. FUCUTUATIONS AND THEIR LINKAGE TO BRANE WORLD PHYSICS

We shall reference a simple Lypunov Exponent argument as to adjustment of the initial quantum flux on the brane world picture. This will next be followed up by a description of how to link the estimated requirement of heat influx needed to get the quantum spatial variation flux in line with inflation expansion parameters.

To begin this, we access the article "Quantum theory without Measurements "to ascertain the role of a Lyapunov exponent $\widetilde{\Lambda}_{chaos}$ such that [38].

$$\Delta p = (\Delta p_0) \cdot \exp(-\widetilde{\Lambda}_{chaos} \cdot t) \tag{8}$$

And

$$\Delta x = (\hbar/\Delta p_0) \cdot \exp(\widetilde{\Lambda}_{chaos} \cdot t) \tag{9}$$

Here, we define where a wave functional forms via the minimum time requirement as to the formation of a wave functional via a minimum time of the order of Planck's time

$$t_h = (\widetilde{\Lambda}_{chaos})^{-1} \cdot \ln[\Delta p_0 \cdot \widehat{L}/\hbar] \approx t_P \tag{10}$$

If we have a specified minimum length as to how to define $\widehat{L} \approx l_P$, this is a good way to get an extremely large $\widetilde{\Lambda}_{chaos}$ value, all in all so that we have Eqn. 3 above on the order of magnitude at the end of inflation as large as what the universe becomes, i.e. a few centimeters or so, from an initial length value on the order of Planck's length $\widehat{L} \approx l_P$.

## IV. WHY WE EVEN BOTHER WITH TALKING ABOUT SUCH A SIMPLIFIED FLUCTUATION PROCEDURE

Two reasons First of all, we have that our description of a link of the sort between a brane world effective potential and Guth style inflation has been partly replicated by Sago, Himenoto, and Sasaki in November 2001 [54] where they assumed a given scalar potential, assuming that *m* is the mass of the bulk scalar field. This permits mixing the false vacuum hypothesis of Coleman [16] in 4 dimensions with brane world theory in five dimensions.

$$V(\phi) = V_0 + \frac{1}{2}m\phi^2 \tag{11}$$

Their model is in part governed by a restriction of their 5-dimensional metric to be of the form, with $l = $ brane world curvature radius, and H their version of the Hubble parameter

$$dS^2 = dr^2 + (H \cdot l)^2 \cdot dS^2{}_{4-\dim} \tag{12}$$

I.e. if we take $k_5^2$ as being a 5 dimensional gravitational constant

$$H = \frac{k_5^2 \cdot V_0}{6} \tag{13}$$

Our difference with Eqn. (11) is that we are proposing that it is an intermediate step, and not a global picture of the inflation field potential system. However, the paper they present with its focus upon the zero mode contributions to vacuum expectations $\langle \delta\phi^2 \rangle$ on a brane has similarities as to what we did which should be investigated further. The difference between what they did, and our approach is in their value of [53]

$$dS^2{}_{4-\dim} \equiv -dt^2 + \frac{1}{H^2} \cdot [\exp(2 \cdot H \cdot t)] \cdot dx^2 \tag{14}$$

This assumes one is still working with a modified Gaussian potential all the way through, as seen in Eqn. (11). This is assuming that there exists an effective five dimensional cosmological parameter which is still less than zero, with $\Lambda_5 < 0$, and $|\Lambda_5| > k_5^2 \cdot V_0$ so that

$$\Lambda_{5,eff} = \Lambda_5 + k_5^2 \cdot V_0 < 0 \qquad (15)$$

It is simply a matter of having

$$|m^2| \cdot \phi^2 << V_0 \qquad (16)$$

And of making the following identification

$$\phi_{5-dim} \propto \tilde{\phi}_{4-dim} \equiv \tilde{\phi} \approx [\phi - \varphi_{fluctuations}]_{4-dim} \qquad (17)$$

With $\varphi_{fluctuations}$ in Eqn. (17) is an equilibrium value of a true vacuum minimum for a chaotic four dimensional quadratic scalar potential for inflationary cosmology. This in the context of the fluctuations having an upper bound of $\tilde{\tilde{\phi}}$ (Here, $\tilde{\tilde{\phi}} \geq \varphi_{fluctuations}$). And $\tilde{\phi}_{4-dim} \equiv \tilde{\tilde{\phi}} - \frac{m}{\sqrt{12 \cdot \pi \cdot G}} \cdot t$, where we use $\tilde{\tilde{\phi}} > \sqrt{\frac{60}{2 \cdot \pi}} M_P \approx 3.1 M_P \equiv 3.1$, with $M_P$ being a Planck mass. This identifies an imbedding structure we will elaborate upon later on. This will in its own way lead us to make sense of a phase transition we will write as a four dimensional embedded structure within the 5 dimensional Sundrum brane world structure and the four dimensional

$$\begin{array}{ll} \tilde{V}_1 & \to \tilde{V}_2 \\ \tilde{\phi}(increase) \leq 2 \cdot \pi & \to \tilde{\phi}(decrease) \leq 2 \cdot \pi \\ t \leq t_P & \to t \geq t_P + \delta \cdot t \end{array} \qquad (18a)$$

The potentials $\tilde{V}_1$, and $\tilde{V}_2$ will be described in terms of **S-S'** di quark pairs [65] nucleating and then contributing to a chaotic inflationary scalar potential system. Here, $m^2 \approx (1/100) \cdot M_P^2$

$$\tilde{V}_1(\phi) \propto \frac{m_a^2(T)}{2} \cdot (1 - \cos(\tilde{\phi})) + \frac{m^2}{2} \cdot (\tilde{\phi} - \phi^*)^2 \qquad (18b)$$

$$\tilde{V}_2(\phi) \propto \frac{1}{2} \cdot (\tilde{\phi} - \phi_C)^2 \qquad (18c)$$

This is where for low temperatures $m_a^2(T \approx \varepsilon^+) \cong 100 \cdot m^2 \xrightarrow{T \to l \arg e} m_a^2(T \approx 10^{12} K) << m^2$ I.e. we look at axion walls specified by Kolb's book [36] about conditions in the early universe (1991) with his Eqn. (10.27) vanishing and collapsing to Guth's quadratic inflation. I.e. having the quadratic contribution to an inflation potential arise due to the vanishing of the axion contribution of the first potential of Eqn. (18a) above with a temperature dependence of

$$V(a) = m_a^2 \cdot (f_{PQ}/N)^2 \cdot (1 - \cos[a/(f_{PQ}/N)]) \qquad (19)$$

Here, he has the mass of the axion potential as given by $m_a$ as well as a discussion of symmetry breaking which occurs with a temperature $T \approx f_{PQ}$. This is done via scaling the axion mass via either [7]

$$m_a(T) \approx 3 \cdot H(T) \tag{20}$$

So that the axion 'matter' will oscillate with a 'frequency' proportional to $m_a(T)$. The hypothesis so presented is that input thermal energy given by the prior universe being inputted into an initial cavity / region dominated by an initially configured low temperature axion domain wall would be thermally excited to reach the regime of temperature excitation permitting an order of magnitude drop of axion density $\rho_a$ from an initial temperature $^2 T_{dS}\big|_{t \leq t_P} \sim H_0 \approx 10^{-33} eV$ as given by( assuming we will use the following symbol $\psi_a$ for axions, and then relate it to Guth inflationary potential scalar fields later on, and state that $\psi_a(t) = \psi_i$ is the initial misaligned value of the field)

$$\rho_a(T_{ds}) \propto \frac{1}{2} \cdot m_a(T_{dS}) \cdot \psi_i^2 \xrightarrow[T \to 10 \ to \ 12th \ power \ Kelvin]{} \varepsilon^+ \tag{21}$$

Or

$$m_a(T) \cong 0.1 \cdot m_a(T=0) \cdot (\Lambda_{QCD}/T)^{3.7} \tag{22}$$

The dissolving of axion walls is necessary for dark matter-dark energy production and we need to incorporate this in a potential system in four dimensions, and relate it to a bigger five dimensional potential systems. We need to find a way to, using brane theory, to investigate how we can have non zero axion mass conditions to begin with. This will be done after we bring up a brief interlude of Quintessence evolution of the scalar field, as presented in **Appendix III**, which for long periods of time is unworkable, but which would be appropriate up to times in the order of magnitude of the Planck's time coefficient. Note after this description of Quintessence, we will be looking at the mechanism of thermal input leading to the time dependence of the axion mass, as given in Eqn. (22) above. We should note that the higher dimensional version of the Bogomolnyi inequality assumed here is covered in [42].

## V. DYNAMICS OF AXION INTERACTION WITH BARYONIC MATTER, VIA QUINTESSENCE SCALAR FIELD

This discussion is modeled on an paper on Quintessence and spontaneous Leptogenesis ( baryogenesis ) by M. Li, X. Wang, B.Feng, and Z. Zhang [43] which gave an effective Lagrangian, and an equation of 'motion' for quintessence which yielded four significant cases for our perusal. The last case, giving a way to reconcile the influx of thermal energy of a quantum bounce into an axion dominated initial cosmology, which lead to dissolution of the excess axion 'mass'. This final reduction of axion 'mass' via temperature variation leads to the Guth style chaotic inflationary regime.

Let us now look at a different effective Lagrangian which has some similarities to equations of motion for Quintessence scalar fields, assuming as was in Eqn. 123 that specifying a non zero value to $(\partial_0 \phi)$ where $(\partial_0 \phi) \neq 0$ is implicitly assumed in Eqn. 18a to Eqn. 18c

i.e. [43]

$$L_{eff} \propto \frac{\tilde{c}}{M} \cdot (\partial_\mu \phi) \cdot J^\mu \qquad (23)$$

What will be significant will be the constant, $\tilde{c}$ which is the strength of interaction between a quintessence scalar field and baryonic matter. M in the denominator is a mass scale which can be either $M \equiv M_{planck}$, or $M \equiv M_{GUT}$ is not so important to our discussion, and $J^\mu$ is in reference to a baryonic 'current'. The main contribution to our analysis this paper gives us is in their quintessence 'equation of motion' which we will present, next. Note, that what we are calling $g_b$ is the degrees of freedom of baryonic states of matter, and $T$ is a back ground temperature w.r.t. early universe conditions. $H \cong 1/t(time)$ is the Hubble parameter, with time $t \propto O(t_P)$, i.e. time on the order of Planck's time, or in some cases much smaller than that.

$$\ddot{\phi} \cdot \left[1 + \frac{\tilde{c}}{M^2} \cdot \frac{T^2}{6} \cdot g_b\right] + 3 \cdot H \cdot \dot{\phi} \cdot \left[1 + \frac{\tilde{c}}{M^2} \cdot \frac{T^2}{6} \cdot g_b\right] + \left(\frac{\partial V_{axion-contri}}{\partial \phi}\right) \cong 0 \qquad (24)$$

Here I am making the following assumption about the axion contribution scalar potential system

$$V_{axion-contri} \equiv f[m_{axion}(T)] \cdot (1 - \cos(\phi)) + \frac{m^2}{2} \cdot (\phi - \phi_C)^2 \qquad (25)$$

For low temperatures, we can assume that prior to inflation, as given by Carroll and Chen we have for $t << t_P$

$$f[m_{axion}(T)]_{T \approx 2^0 Kelvin} \propto O((50-100) \cdot m^2) \qquad (26)$$

And that right at the point where we have a thermal input with back ground temperatures at or greater than $10^{12} Kelvin$ we are observing for $0 < \varepsilon^+ << 1$ and times $t << t_P$

$$f[m_{axion}(T)]_{T \approx 10^{12} Kelvin} \propto O((\varepsilon^+) \cdot m^2) \qquad (27)$$

This entails having at high enough temperatures

$$V_{axion-contri}\big|_{T > 10^{12} Kelvin} \cong \frac{m^2}{2} \cdot (\phi - \phi_C)^2 \qquad (28)$$

Those who are interested in the short lived Quintessence phenomena involved should look at **Appendix III** which in the fourth case re enforces the existence of conditions for which Eqn 28 arises. We will refer to this later on in this paper. Needless to say, we have that the upshot is, that for large, but shrinking axion mass contributions we have a cyclical oscillatory system, which breaks down and becomes a real field if the axion mass disappears. First of all, though, we have to understand how the conditions presented by S. Carroll, and J. Chen [15] came about via brane theory. Our point is that we need brane theory to establish the initial starting point of reference for low temperature conditions at the onset of inflation.

# VI. SETTING CONDITIONS FOR LOW TEMPERATURE AND LOW ENTROPY BOUNDS VIA BRANE WORLD PHYSICS FOR THE START OF INFLATION

Our starting point here is first showing equivalence of entropy formulations in both the Brane world and the more typical four dimensional systems. A Randall-Sundrum Brane world will have the following as a line element and we will continue from here to discuss how it relates to holographic upper bounds to both anti De sitter metric entropy expressions and the physics of dark energy generating systems.

To begin with, let us first start with the following as a $A \cdot dS_5$ model of tension on brane systems, and the line elements. If there exists a tension $\breve{T}$, with Plank mass in five dimensions denoted as $M_5$, and a curvature value of $l$ on $A \cdot dS_5$ we can write [46]

$$\breve{T} = 3 \cdot \left( M_5^3 / 4 \cdot \pi \cdot l \right) \tag{29}$$

Furthermore, the $A \cdot dS_5$ line element, with $r =$ distance from the brane, becomes [14]

$$\frac{dS^2}{l^2} = \left( \exp(2 \cdot r) \right) \cdot \left[ -dt^2 + d\rho^2 + \sin^2 \rho \cdot d\Omega_2 \right] + dr^2 \tag{30}$$

We can then speak of a four dimensional volume $V_4$ and its relationship with a three dimensional volume $V_3$ via

$$V_4 = l \cdot V_3 \tag{31}$$

And if a Brane world gravitational constant expression $G_N = M_4^{-2} \Leftrightarrow M_4^2 = M_5^3 \cdot l$ we can get a the following space bound Holographic upper bound to entropy [14]

$$S_5(V_4) \leq V_3 \cdot \left( M_5^3 / 4 \right) \tag{32}$$

If we look at an area 'boundary' $A_2$ for a three dimensional volume $V_3$, we can re cast the above holographic principle to (for a volume $V_3$ in Planck units)

$$S_4(V_3) \leq A_2 \cdot \left( M_4^2 / 4 \right) \tag{33}$$

We link this to the principle of the Jeans inequality for gravitational physics and a bound to entropy and early universe conditions, as given by S. Carroll and J.Chen (2005) [15] by stating if $S_4(V) = S_5(V_4)$ then if we can have

$$A_2 \xrightarrow[t \to t_P]{} \varepsilon_{small\ area} \Leftrightarrow S_5(V_4) \approx \delta_{small\ entropy} \tag{34}$$

Low entropy conditions for initial conditions, as stated above give a clue as to the likely hood of low temperatures as a starting point via R. Easther et al. (1998) relationship of a generalized non brane world entropy bound, assuming that $n^* \approx$ bosonic degrees of freedom and $T$ as generalized temperature, so we have as a temperature based elaboration of the original work by Susskind on holographic projections forming area bound values to entropy

$$\frac{S}{A} \leq \sqrt{n^*} \cdot T \tag{35}$$

Similar reasoning, albeit from the stand point of the Jeans inequality and instability criteria lead to Sean Carroll and J. Chen (2005) giving for times at or earlier than the Planck time $t_P$ that a vacuum state would initially start off with a very low temperature

$$T_{dS}\big|_{t \leq t_P} \sim H_0 \approx 10^{-33} eV \tag{36}$$

We shall next refer to how this relates to, considering a low entropy system as a start an expression Wheeler wrote for graviton production and its implications for early relic graviton production, and its connection to axion walls and how they subsequently vanish at or slightly past the Planck time $t_P$. This involves using relic graviton production as referenced next as a way to mark the influx of thermal heat. We after talking about thermal input will integrate the following equation to obtain a phenomenological way to mark how gravitons may arise naturally.

## VII. WHEELER GRAVITON PRODUCTION FORMULA FOR RELIC GRAVITONS

As is well known, a good statement about the number of gravitons per unit volume with frequencies between $\omega$ and $\omega + d\omega$ may be given by (assuming here, that $\bar{k} = 1.38 \times 10^{-16} erg/^0K$, and $^0K$ is denoting Kelvin temperatures, while we keep in mind that Gravitons have two independent polarization states), as given by Weinberg [62]

$$n(\omega)d\omega = \frac{\omega^2 d\omega}{\pi^2} \cdot \left[\exp\left(\frac{2 \cdot \pi \cdot \hbar \cdot \omega}{\bar{k}T}\right) - 1\right]^{-1} \tag{37}$$

This formula predicts what was suggested earlier. A surge of gravitons commences due to a rapid change of temperature. I.e. if the original temperature were low, and then the temperature rapidly would heat up? Here is how we can build up a scenario for just that. Eqn. (37) suggests that at low temperatures we have large busts of gravitons.

Now, how do we get a way to get the $\omega$ and $\omega + d\omega$ frequency range for gravitons, especially if they are relic gravitons? First of all, we need to consider that certain researchers claim that gravitons are not necessarily massless, and in fact the Friedman equation acquires an extra dark-energy component leading to accelerated expansion. The mass of the graviton allegedly can be as large as $\sim (10^{15} cm)^{-1}$. This is though if we connect massive gravitons with dark matter candidates, and not necessarily with relic gravitons. Having said this we can note that Massimo Giovannini writes in an introduction to his Phys Rev D article [25] about presenting a model which leads to post-inflationary phases whose effective equation of state is stiffer than radiation. He states: *The expected gravitational wave logarithmic energy spectra are tilted towards high frequencies and characterized by two parameters: the inflationary curvature scale at which the transition to the stiff phase occurs and the number of (nonconformally coupled) scalar degrees of freedom whose decay into fermions triggers the onset of a gravitational reheating of the Universe. Depending upon the parameters of the model and upon the different inflationary dynamics (prior to the onset of the stiff evolution), the relic gravitons energy density can be much more sizable than in standard inflationary models, for frequencies larger than 1 Hz.* Giovannini [25] claims that there are grounds for an energy density of relic gravitons in critical units (i.e., $h_0^2 \Omega_{GW}$) is of the order of $10^{-6}$, roughly eight orders of magnitude larger than in ordinary inflationary models. That roughly corresponds with what could be expected in our brane world model for relic graviton production.

We are using a different value for the frequency, namely the one given by threshold energy, Eqn. 7a, divided by $\hbar$, but we reference Giovannini's article [25] because it also predicts a far greater surge of relic gravitons than is usually mentioned in the literature. Still though, in its own way, this frequency range is in line with the energy value we can read off Weinberg's threshold value of energy, divided by Planck's constant, of a starting point for genuine quantum effects in quantum gravity models we refer to in Eqn. (7a) above [62]

We also are as stated earlier, stating that the energy input into the frequency range so delineated comes from a prior universe collapse, as modeled by Ashtekar, A., Pawlowski, T. and Singh, P (2006) [5,6] via their quantum bounce model as given by quantum loop gravity calculations. We will state more about this later in this document.

. The hypothesis so presented is that input thermal energy given by the prior universe being inputted into an initial cavity / region dominated by an initially configured low temperature axion domain wall would be thermally excited to reach the regime of temperature excitation permitting an order of magnitude drop of axion density $\rho_a$ from an initial temperature $T_{dS}\big|_{t \leq t_P} \sim H_0 \approx 10^{-33} eV$ [15]. . We shall, before doing this reference graviton production

## VIII. GRAVITON SPACE PROPULSION SYSTEMS

We need to understand what is required for realistic space propulsion. To do this, we need to refer to a power spectrum value which can be associated with the emission of a graviton. Fortunately, the literature contains a working expression as to power generation for a graviton being produced for a rod spinning at a frequency per second $\omega$, which is by Fontana (2005) at a STAIF new frontiers meeting, which allegedly gives for a rod of length $\hat{L}$ and of mass m a formula for graviton production power, [24]

$$P(power) = 2 \cdot \frac{m_{graviton}^2 \cdot \hat{L}^4 \cdot \omega_{net}^6}{45 \cdot (c^5 \cdot G)} \tag{38}$$

The point is though that we need to say something about the contribution of frequency needs to be understood as a mechanical analogue to the brute mechanics of graviton production. For the sake of understanding this, we can view the frequency $\omega_{net}$ as an input from an energy value, with graviton production number (in terms of energy) as given approximately via an integration of Eqn. (37) above, $\hat{L} \propto l_P$, mass $m_{graviton} \propto 10^{-60} kg$. It also depends upon a HUGE **number** of relic gravitons being produced, due to the temperature variation so proposed. We can see the results of integrating Eqn. (37)

$$\langle n(\omega) \rangle = \frac{1}{\omega(net\ value)} \int_{\omega 1}^{\omega 2} \frac{\omega^2 d\omega}{\pi^2} \cdot \left[ \exp\left(\frac{2 \cdot \pi \cdot \hbar \cdot \omega}{\bar{k}T}\right) - 1 \right]^{-1} \tag{39}$$

And then one can set a normalized 'energy input 'as $E_{eff} \equiv \langle n(\omega) \rangle \cdot \omega \equiv \omega_{eff}$; with $\hbar \omega \xrightarrow[\hbar \equiv 1]{} \omega \equiv |E_{critical}|$ being given in Eqn. (7a) above, which leads to the following table of results, with $T^*$ being an initial temperature of the pre inflationary universe condition

### HOW TO OUTLINE THE EXISTENCE OF A RELIC GRAVITON BURST

| N1=1.794 E-6 for $Temp = T^*$ | Power = 0 |
|---|---|
| N2=1.133 E-4 for $Temp = 2T^*$ | Power = 0 |
| N3= 7.872 E+21 for $Temp = 3T^*$ | Power = 1.058 E+16 |
| N4= 3.612E+16 for $Temp = 4T^*$ | Power $\cong$ very small value |
| N5= 4.205E-3 for $Temp = 5T^*$ | Power= 0 |

The outcome is that there is a distinct power spike associated with Eqn. 38 and Eqn. 39, which is congruent with a relic graviton burst, assuming when one does this that the back ground in the initial inflation state causes a thermal heat up of the axion wall 'material' due to a thermal input from a prior universe quantum bounce. Our next task will be to configure the conditions via brane world dynamics leading to graviton production. This necessitates using a brane world potential to accommodate the building of a structure accommodating a transition from relic graviton production to the onset of Guth style chaotic inflation.

## IX. RANDALL SUNDRUM EFFECTIVE POTENTIAL

The consequences of the fifth-dimension show up in a simple warped compactification involving two branes, i.e., a Planck world brane, and an IR brane. Let's call the brane where gravity is localized the Planck brane. This construction permits (assuming K is a constant picked to fit brane world requirements) [55]

$$S_5 = \int d^4x \cdot \int_{-\pi}^{\pi} d\theta \cdot R \cdot \left\{ \frac{1}{2} \cdot (\partial_M \phi)^2 - \frac{m_5^2}{2} \cdot \phi^2 - K \cdot \phi \cdot [\delta(x_5) + \delta(x_5 - \pi \cdot R)] \right\} \tag{40}$$

Here, what is called $m_5^2$ can be linked to Kaluza Klein "excitations" via (for a number $n > 0$)

$$m_n^2 \equiv \frac{n^2}{R^2} + m_5^2 \tag{41}$$

To build the Kaluza–Klein theory, one picks an invariant metric on the circle $S^1$ that is the fiber of the $U(1)$-bundle of electromagnetism. This leads to construction of a two component scalar term with contributions of different signs. i.e.

$$S_5 = -\int d^4x \cdot V_{eff}(R_{phys}(x)) \to -\int d^4x \cdot \tilde{V}_{eff}(R_{phys}(x)) \tag{42}$$

We should briefly note what an effective potential is in this situation.

We get

$$\tilde{V}_{eff}(R_{phys}(x)) = \frac{K^2}{2 \cdot m_5} \cdot \frac{1 + \exp(m_5 \cdot \pi \cdot R_{phys}(x))}{1 - \exp(m_5 \cdot \pi \cdot R_{phys}(x))} + \frac{\tilde{K}^2}{2 \cdot \tilde{m}_5} \cdot \frac{1 - \exp(\tilde{m}_5 \cdot \pi \cdot R_{phys}(x))}{1 + \exp(\tilde{m}_5 \cdot \pi \cdot R_{phys}(x))} \tag{43}$$

This above system has a metastable vacuum for a given special value of $R_{phys}(x)$. Start with [64]

$$\Psi \propto \exp(-\int d^3x_{space} d\tau_{Euclidian} L_E) \equiv \exp(-\int d^4x \cdot L_E) \tag{44}$$

$$L_E \geq |Q| + \frac{1}{2} \cdot (\tilde{\phi} - \phi_0)^2 \{ \} \xrightarrow{Q \to 0} \frac{1}{2} \cdot (\tilde{\phi} - \phi_0)^2 \cdot \{ \} \tag{45}$$

Part of the integrand in Eqn. (44) is known as an action integral, $S = \int L\, dt$, where L is the Lagrangian of the system. Where as we also are assuming a change to what is known as Euclidean time, via $\tau = i \cdot t$, which has the effect of inverting the potential to emphasize the quantum bounce hypothesis of Sidney Coleman. In that hypothesis, $L$ is the Lagrangian with a vanishing kinetic energy contribution, i.e. $L \to V$, where $V$ is a potential whose graph is 'inverted' by the Euclidian time. Here, the spatial dimension $R_{phys}(x)$ is defined so that

$$\tilde{V}_{eff}(R_{phys}(x)) \approx \text{constant} + \tfrac{1}{2} \cdot (R_{phys}(x) - R_{critical})^2 \propto \tilde{V}_2(\tilde{\phi}) \propto \frac{1}{2} \cdot (\tilde{\phi} - \phi_C)^2 \tag{46}$$

And

$$\{\ \} = 2 \cdot \Delta \cdot E_{gap} \tag{47}$$

We should note that the quantity $\{\ \} = 2 \cdot \Delta \cdot E_{gap}$ referred to above has a shift in minimum energy values between a false vacuum minimum energy value, $E_{\text{false min}}$, and a true vacuum minimum energy $E_{\text{true min}}$, with the difference in energy reflected in Eqn. (47) above.

This requires, if we take this analogy seriously the following identification with what was done by the Japanese theorists

$$\tilde{V}_{eff}(R_{phys}(x)) \approx \mathbf{Constant} + \frac{1}{2} \cdot (R_{phys}(x) - R_{critical})^2 \propto V_0 + \frac{m}{2} \cdot [\phi - \varphi_{fluctuations}]^2_{4-\dim} \tag{48}$$

So that one can make equivalence between the following statements. These need to be verified via serious analysis.

$$\mathbf{Constant} \leftrightarrow V_0 \tag{48a}$$

$$\frac{1}{2} \cdot (R_{phys}(x) - R_{critical})^2 \leftrightarrow \frac{m}{2} \cdot [\phi - \varphi_{fluctuations}]^2_{4-\dim} \tag{48b}$$

## X. USING OUR BOUND TO THE COSMOLOGICAL CONSTANT

We use our bound to the cosmological constant to obtain a conditional escape of gravitons from an early universe brane. To begin, we present conditions (Leach and Lesame, 2005) [40] for gravitation production. Here $R$ is proportional to the scale factor 'distance'.

$$B^2(R) = \frac{f_k(R)}{R^2} \tag{49}$$

Also there exists an 'impact parameter'

$$b^2 = \frac{E^2}{P^2} \tag{50}$$

This leads to, practically, a condition of 'accessibility' via R so defined with respect to 'bulk dimensions'

$$b \geq B(R) \tag{51}$$

$$f_k(R) = k + \frac{R^2}{l^2} - \frac{\mu}{R^2} \tag{52}$$

Here, k = 0 for flat space, k = -1 for hyperbolic three space, and k = 1 for a three sphere, while an radius of curvature

$$l \equiv \sqrt{\frac{-6}{\Lambda_{5-\dim}}} \tag{53}$$

This assumes a negative bulk cosmological constant $\Lambda_{5-\dim}$ and that $\mu$ is a five-dimensional Schwartz shield mass. We assume emission of a graviton from a bulk horizon via scale factor, so $R_b(t) = a(t)$. Then we have a maximum effective potential of gravitons defined via

$$B^2(R_t) = \frac{1}{l^2} + \frac{1}{4 \cdot \mu} \tag{54}$$

This leads to a bound with respect to release of a graviton from an anti De Sitter brane [40] as

$$b \geq B(R_t) \tag{55}$$

In the language of general relativity, anti de Sitter space is the maximally symmetric, vacuum solution of Einstein's field equation with a negative cosmological constant Λ.

How do we link this to our problem with respect to di quark contributions to a cosmological constant? Here we make several claims.

**Claim 1**: It is possible to redefine $l \equiv \sqrt{-6/\Lambda_{5-dim}}$ as

$$l_{eff} = \sqrt{\left|\frac{6}{\Lambda_{eff}}\right|} \tag{56}$$

**Proof of Claim 1**: There is a way, for finite temperatures for defining a given four-dimensional cosmological constant (Park, Kim,) [50]. Check Appendix **II** below for the steps while noting that the end result is that

$$\Lambda_{5-dim} \xrightarrow[external\ temperature \to small]{} \text{Large value} \tag{57}$$

And set

$$|\Lambda_{5-dim}| = \Lambda_{eff} \tag{58}$$

In working with these values, we should pay attention to how $\cdot\Lambda_{4-dim}$ is defined by Park, et al. [50] with $\varepsilon* = \frac{U_T^4}{k^*}$ and $U_T \propto (external\ temperature)$, and $k^* = \left(\frac{1}{'AdS\ curvature}\right)$ so that

$$\cdot\Lambda_{4-dim} = 8 \cdot M_5^3 \cdot k^* \cdot \varepsilon* \xrightarrow[external\ temperature \to 3 Kelvin]{} (.0004eV)^4 \tag{59}$$

Here, we define $\Lambda_{eff}$ as being an input from Eqn. (18a) to (18b) to Eqn (18c) due to, in part

$$\Delta\Lambda_{total}\big|_{effective} = \lambda_{other} + \Delta V$$
$$\xrightarrow[\Delta V \to end\ chaotic\ inflation\ potential]{} \Lambda_{observed} \cong \Lambda_{4-dim}(3\,Kelvin) \tag{60}$$

This, for potential $V_{min}$, is defined via transition between the first and the second potentials of Eqn. (18b) and Eqn. (18c)

$$B_{eff}^{\;2}(R_t) = \frac{1}{l_{eff}^{\;2}} + \frac{1}{4 \cdot \mu} \tag{61}$$

**Claim 2**: $R_b(t) = a(t)$ ceases to be definable for times where the upper bound to the time limit is in terms of Planck time and in fact the entire idea of a de Sitter metric is not definable in such a physical regime. This is a given in standard inflationary cosmology where traditionally the scale factor in cosmology is a, parameter of the Friedmann-Lemaître-Robertson-Walker model, and is a function of time which represents the relative expansion of the universe. It relates physical coordinates (also called proper coordinates) to co moving coordinates. For the FLRW model

$$L = \bar{\bar{\lambda}} \cdot a(t) \tag{62}$$

Where L is the physical distance $\overline{\tilde{\lambda}}$ is the distance in co moving units, and $a(t)$ is the scale factor. While general relativity allows one to formulate the laws of physics using arbitrary coordinates, some coordinate choices are natural choices, which are easy to work with. *Comoving coordinates* are an example of such a natural coordinate choice. They assign constant spatial coordinate values to observers who perceive the universe as isotropic. Such observers are called *comoving observers* because they move along with the Hubble flow. *Comoving distance* is the distance between two points measured along a path of constant cosmological time. It can be computed by using $t_e$ as the lower limit of integration as a time of emission

$$\overline{\tilde{\lambda}} \equiv \int_{t_e}^{t} \frac{c \cdot dt'}{a(t')} \tag{63}$$

This claim 2 breaks down completely when one has a strongly curved space, which is what we would expect in the first instant of less than Planck time evolution of the nucleation of a new universe.

**Claim 3**: Eqn. (60) has a first potential which tends to be for a di quark nucleation procedure which just before a defined Planck's time $t_P$. But that the cosmological constant was prior to time $t_P$ likely far higher, perhaps in between the values of the observed cosmological constant of today, and the QCD tabulated cosmological constant which is $10^{120}$ time greater. i.e.

$$b^2 \geq B_{eff}^{\ 2}(R_t) = \frac{1}{l_{eff}^{\ 2}} + \frac{1}{4 \cdot \mu} \tag{64}$$

Which furthermore

$$\left. \frac{1}{l_{eff}^{\ 2}} \right|_{t \leq t_P} >> \left. \frac{1}{l_{eff}^{\ 2}} \right|_{t \equiv t_P + \Delta(time)} \tag{65}$$

So then that there would be a great release of gravitons at or about time $t_P$.

**Claim 4**: Few gravitons would be produced significantly after time $t_P$.

**Proof of Claim 4**: This comes as a result of temperature changes after the initiation of inflation and changes in value of

$$\left( \Delta l_{eff} \right)^{-1} = \left( \sqrt{\left| \frac{6}{\Lambda_{eff}} \right|} \right)^{-1} \propto \Delta \left( external \quad temperature \right) \tag{66}$$

After this, we need to discuss how this thermal input into the axion wall occurs, leading to these results.

## XI. DI QUARK POTENTIAL SYSTEMS AND THE WHEELER DE-WITT EQUATION

Abbay Ashtekar's quantum bounce gives a discrete version of the Wheeler De Witt equation, [5, 6]. To give us a flavor of what is coming we begin with Henrique's more intuitive approach, as given below [33]

$$\psi_\mu(\phi) \equiv \psi_\mu \cdot \exp(\alpha_\mu \cdot \phi^2) \tag{67}$$

As well as energy term

$$E_\mu = \sqrt{A_\mu \cdot B_\mu} \cdot m \cdot \hbar \tag{68}$$

$$\alpha_\mu = \sqrt{B_\mu / A_\mu} \cdot m \cdot \hbar \tag{69}$$

This is for a 'cosmic' Schrodinger equation as given by

$$\tilde{\hat{H}} \cdot \psi_\mu(\phi) = E_\mu(\phi) \tag{70}$$

This has $V_\mu$ is the Eigen value of a so called volume operator. So:

$$A_\mu = \frac{4 \cdot m_{pl}}{9 \cdot l_{pl}^9} \cdot \left(V_{\mu+\mu_0}^{1/2} - V_{\mu-\mu_0}^{1/2}\right)^6 \tag{71}$$

And

$$B_\mu = \frac{m_{pl}}{l_{pl}^3} \cdot (V_\mu) \tag{72}$$

Ashtekar works [5,6] with as a simplistic structure with a revision of the differential equation assumed in Wheeler-De Witt theory to a form characterized by $\partial^2/\partial\phi^2 \cdot \Psi \equiv -\Theta \cdot \Psi$, and $\Theta \neq \Theta(\phi)$. This will lead to $\Psi$ having roughly the form alluded to in Eqn. (67), which in early universe geometry will eventually no longer be $L^P$, but will have a discrete geometry. This may permit an early universe 'quantum bounce' and an outline of an earlier universe collapsing, and then being recycled to match present day inflationary expansion parameters. The main idea behind the quantum theory of a (big) quantum bounce is that, as density approaches infinity, so the behavior of the quantum foam changes. The foam is a qualitative description of the turbulence that the phenomenon creates at extremely small distances of the order of the Planck length. Here $V_\mu$ is the Eigen value of a so called volume operator and we need to keep in mid that the main point made above, is that a potential operator based upon a quadratic term leads to a Gaussian wave function with an exponential similarly dependent upon a quadratic $\phi^2$ exponent. To get to this, we need to consider a discrete wave functional for a modified Wheeler de Witt equation we would write up as follows. The point is that if we understand the contribution of Eqn. (58) and Eqn. (59) above to space time dynamics, we will be able to confirm or falsify the existence of space time conditions as given by a non $L^P$ structure as implied below. This will entail either confirming or falsifying the structure given to $\Theta$. Also, and more importantly the above mentioned $\Theta$ is a difference operator, allowing for a treatment of the scalar field as an 'emergent time', or 'internal time' so that one can set up a wave functional built about a Gaussian wave functional defined via

$$\max \tilde{\Psi}(k) = \tilde{\Psi}(k)\big|_{k \equiv k^*} \tag{73}$$

This is for a crucial 'momentum' value

$$p_\phi^* = -\left(\sqrt{16 \cdot \pi \cdot G \cdot \hbar^2 / 3}\right) \cdot k^* \tag{74}$$

And

$$\phi^* = -\sqrt{3/16 \cdot \pi G} \cdot \ln|\mu^*| + \phi_0 \tag{75}$$

Which leads to, for an initial point in 'trajectory space' given by the following relation $(\mu^*, \phi_0)$ = (initial degrees of freedom [dimensionless number] ~'Eigen value of 'momentum', initial 'emergent time ' ) so that if we consider Eigen functions of the De Witt (difference) operator, as contributing toward

$$e_k^s(\mu) = (1/\sqrt{2}) \cdot [e_k(\mu) + e_k(-\mu)] \tag{76}$$

With each $e_k(\mu)$ an Eigen function of $\Theta$ above, we have a potentially numerically treatable early universe wave functional data set which can be written as [5, 6]

$$\Psi(\mu,\phi) = \int_{-\infty}^{\infty} dk \cdot \tilde{\Psi}(k) \cdot e_k^s(\mu) \cdot \exp[i\omega(k)\cdot\phi] \qquad (77)$$

The existence of gravitons in itself would be able to either confirm or falsify the existence of non $L^P$ structure in the early universe. This structure was seen as crucial to Ashtekar, A, Pawlowski, T. and Singh, in their arXIV article [5, 6] make reference to a revision of this momentum operation along the lines of basis vectors $|\mu\rangle$ by

$$\hat{p}_t |\mu\rangle = \frac{8\cdot\pi\cdot\gamma\cdot l_{PL}^2}{6}\cdot\mu|\mu\rangle \qquad (78)$$

With the advent of this re definition of momentum we are seeing what Ashtekar works with as a simplistic structure with a revision of the differential equation assumed in Wheeler – De Witt theory to a form characterized by [5,6]

$$\frac{\partial^2}{\partial\phi^2}\cdot\Psi \equiv -\;\Theta\cdot\Psi \qquad (79)$$

$\Theta$ in this situation is such that

$$\Theta \neq \Theta(\phi) \qquad (80)$$

This in itself would permit confirmation of if or not a quantum bounce condition existed in early universe geometry, according to what Ashtekar's two articles predict. In addition it also corrects for another problem. Prior to brane theory we had a too crude model. Why? When we assume that a radius of an early universe—assuming setting the speed of light $c \equiv 1$ is of the order of magnitude $3\cdot(\Delta t \cong t_P)$—we face a rapidly changing volume that is heavily dependent upon a first order phase transition, as affected by a change in the degrees of freedom given by $\cdot(\Delta N(T))_P$. Without gravitons and brane world structure, such a model is insufficient to account for dark matter production and fails to even account for Baryogenesis. It also will lead to new graviton detection equipment re configuration well beyond the scope of falsifiable models configured along the lines of simple phase transitions given for spatial volumes (assuming c = 1) of the form [8]

$$\Delta t \cong t_P \propto \frac{1}{4\pi}\cdot\sqrt{\frac{45}{\pi\cdot(\Delta N(T))_P}}\cdot\left(\frac{M_p}{T^2}\right) \qquad (81)$$

This creates problems, so we look for other ways to get what we want. Grushchuk writes that the energy density of relic gravitons is expressible as [27]

$$\varepsilon(v) \equiv \frac{\pi}{(2\cdot\pi)^4}\cdot\frac{1}{a(t)^4}\cdot H_i^2\cdot H_f^2\cdot a(t)_f^4 \qquad (82)$$

Where the subscripts $i$ and $f$ refer to initial and final states of the scale factor, and Hubble parameter. This expression though is meaningless in situations when we do not have enough data to define either the scale factor, or Hubble parameter at the onset of inflation. How can we tie in with the Gaussian wave functional $\tilde{\Psi}(k)$ defined as an input into the data used to specify Ashtekar's quantum bounce? Here, we look at appropriate choices for an optimum momentum value for specifying a high level of graviton production. If gravitons are, indeed, for dark energy, as opposed to dark matter, without mass, we can use, to first approximation something similar to using the zeroth component of momentum $p^0 = E(energy)/c$, calling $E(energy) \equiv \varepsilon(v)\cdot$(initial nucleation volume), we can read off from Eqn. (67), 'pre inflationary' universe values for the k values of Eqn. (81) can be obtained, with an optimal value selected. This is equivalent to using to first approximation the following. The absolute value of $k^*$, which we call $|k^*|$ is

$$|k^*| = \sqrt{3/16 \cdot \pi \cdot G \cdot \hbar^2} \cdot \left( \varepsilon(v) \cdot \left( initial \quad nucleation \quad volume \right) / c \right) \tag{83}$$

An appropriate value for a Gaussian representation of an instanton awaits more detailed study. But for whatever it is worth we can refer to the known spaleraton value for a multi dimensional instanton via the following procedure. We wish to have a finite time for the emergence of this instanton from a pre inflation state.

If we have this, we are well on our way toward fixing a range of values for $\omega 2 < \omega(net) < \omega 1$, which in turn will help us define

$$\varepsilon(v) \cdot \left( initial \quad volume \right) \approx \hbar \cdot \omega(net) \equiv p^* \cdot c \tag{84}$$

in order to get a value for $k^*$. This value for $k^*$ can then is used to construct a Gaussian wave functional about $k^*$ of the form, as an anzatz. To put into Eqn. (77) above.

$$\Psi(k) \approx \frac{1}{Value} \cdot \exp\left( - c_2 \cdot (k - k^*)^2 \right) \tag{85}$$

If so, then, most likely, the question we need to ask though is the temperature of the 'pre inflationary' universe and its link to graviton production. This will be because the relic graviton production would be occurring before the nucleation of a scalar field. We claim, as beforehand that this temperature would be initially quite low, as given by the two University of Chicago articles, but then rising to a value at or near $10^{12}$ degrees Kelvin after the dissolving of the axion wall contribution given in the dominant value of Eqn. (18b) leading to Eqn (18c) for a chaotic inflationary potential. And now we shall consider why we need to look at relic graviton production, any way.

## XII. DETECTION OF GRAVITONS AS SPIN 2 OBJECTS VIA AVAILABLE DETECTOR SYSTEMS?

To briefly review what we can say now about standard graviton detection schemes, as mentioned above, Rothman wrote Dyson doubts we will be able to detect gravitons via present detector technology. The conundrum is that if one defines the criterion for observing a graviton as [54]

$$\frac{f_\gamma \cdot \sigma}{4 \cdot \pi} \cdot \left( \frac{\alpha}{\alpha_g} \right)^{3/2} \cdot \frac{M_s}{R^2} \cdot \frac{1}{\varepsilon_\gamma} \geq 1 \tag{86}$$

Here,

$$f_\gamma = \frac{L_\gamma}{L} \tag{87}$$

This has $f_\gamma \approx \breve{L}_\gamma / \breve{L}$ as a graviton sources luminosity divided by total luminosity and $R$ as the distance from the graviton source, to a detector. Furthermore, $\alpha = e^2/\hbar$ and $\alpha_g = Gm_p^2/\hbar$ a constant while $\varepsilon_\gamma$ is the graviton potential energy. Here, $\breve{L}_\gamma$ – luminosity of graviton producing process $\geq 7.9$ x 10 to the 14th ergs/s, while $\breve{L}$ – general background luminosity which is $usually \gg \breve{L}_\gamma$. At best, we usually can set $f_\gamma = .02$, which does not help us very much. That means we need to look else where than the usual processes to get satisfaction for Graviton detection. This in part is why we are looking at relic graviton production for early universe models, usually detectable via the criteria developed for white dwarf stars of one graviton for $10^{13-14}$ neutrinos [56].

We should state that we will generally be referring to a cross section which is frequently the size of the square of Planck's length $l_P$ which means we really have problems in detection, if the luminosity is so low. An upper bound to the cross section $\sigma$ for a graviton production process $\approx 1/M$ with $M - Planck$ scale in 4+n dimensions $\equiv (M_P^2/\hat{V}_n)^{1/2+n}$, and this is using a very small $\hat{V}_n - Compactified$ early universe extra dimension 'square' volume $\approx$ 10-15 mm per side.

.As stated in the manuscript, the problem then becomes determining a cross section $\sigma$ for a graviton production process and $f_\gamma = L_\gamma / L$. Here, a 4-dimensional graviton emission cross section goes like 1/M. The existence of branes is relevant to graviton production. And understanding the link between the brane world prediction of relic graviton production and Wheelers numerical model is crucial to understanding more of the origins of the quantum geometry problem of early universe conditions.

## XIII. CONCLUSION

So far, we have tried to reconcile the following.

First is that Brane world models will not permit Akshenkar's quantum bounce [5, 6]. The quantum bounce idea is used to indicate how one can reconcile axion physics with the production of dark matter/dark energy later on in the evolution of the inflationary era where one sees Guth style chaotic inflation for times $t \geq t_P$ and the emergence of dark energy during the inflation era.

In addition is the matter of Sean Carroll, J. Chens paper which pre supposes a low entropy – low temperature pre inflationary state of matter prior to the big bang. How does one ramp up to the high energy values greater than temperatures $10^{12}$ Kelvin during nucleosynthesis? The solution offered is novel and deserves further inquiry and investigation.

Thirdly is the issue of relic graviton production. How to observe it? The last section about limitations of graviton detectors, as opposed to gravity waves points to the obvious problems. Current estimates speak of a detector system the size of Jupiter to be able to detect a single graviton. This is patently absurd and needs to be addressed, likely by coupling graviton production with neutrino physics

We can point to the following as tentative successes of our model which need further elaboration

Gravitons would appear to be produced in great number in the $\Delta t \approx t_P$ neighborhood, according to a brane world interpretation just given. This depends upon the temperature dependence of the 'cosmological constant.' And is for a critical temperature $T_C$ defined in the neighborhood of an initial grid of time $\Delta t \approx t_P$. We need to reverse engineer conditions which would give a chance of using indirect ways to observe this sort of graviton production, via methods similar to what S.L. Shapiro, and S. Teukolosky, wrote up in "Black holes, white dwarfs, and Neutron stars", John Wiley and Sons, New York, New York, 1983, where they fixed a ratio between gravitons and neutrinos.

A Randall-Sundrum effective potential, as outlined herein, would give a structure for embedding an earlier than axion potential structure, which would be a primary candidate for an initial configuration of dark energy .This structure would, by baryogenesis, be a shift to dark energy. We need to get JDEM space observations configured to determine if WIMPS are in any way tied into the supposed dark energy released after a $\Delta t \approx t_P$ time interval.

In doing this, we should note the following. We have reference multiple reasons for an initial burst of graviton activity, i.e. if we wish to answer Freeman Dyson's question about the existence of gravitons in a relic graviton stand point.

Now for suggestions as to future research. We are in this situation making reference to solving the cosmological "constant" problem without using G. Gurzadyan and She-Sheng Xue's approach which is fixed upon the scale factor $a(t)$ for a present value of the cosmological constant. We wish to obtain, via Parks method of linking four and five dimensional cosmological constants a way to obtain a temperature based initial set of conditions for this parameter, which would eliminate the need for the scale factor being appealed to, all together. In doing so we also will attempt to either confirm or falsify via either observations from CMB based systems, or direct neutrino physics counting of relic graviton production the exotic suggestions given by Holland and Wald for pre inflation physics and/or shed light as to the feasibility of some of the mathematical suggestions given for setting the cosmological constant parameter given by other researchers.

As I was asked about earlier, this does have a directional component which was given by Weinberg in 1972 as the power per solid angle [62]

$$\frac{dP}{d\Omega} = \frac{G \cdot \omega_{net}^6}{4 \cdot \pi} \cdot \Lambda_{ij,lm}(k) \cdot D_{ij}^*(\omega) \cdot D_{lm}(\omega) \tag{88}$$

Where we can write $D_{ij}(\omega)$ in terms of a Fourier transform of the $T_{ij}(x,\omega)$ energy- momentum tensor

$$D_{ij}(\omega) \equiv -\frac{2}{\omega^2} \cdot T_{ij}(k,\omega) \tag{89}$$

Getting realistic values of $T_{ij}(x,\omega)$ will entail a lot of work and will entail trying to make sense of curvature conditions at the onset of inflation which are currently glossed over. If we can make sense of this, we will be fulfilling the aims of the loop quantum gravity researchers who are attempting to understand early universe geometry at the point of quantum singularities. One other datum is that the width of a graviton is close to Planck length in dimensional considerations. About 100,000 times smaller than the value of the Proton itself, and near the value of a 'sphere' itself for implementation of Eqn. (88) above. That suggests very simple ways to implement curvature considerations, if we can come up with direct calculation of Eqn. (89) above. We will be making full use of such simplifications in a subsequent numerical evaluation of Eqn. (88) above.

Doing all of this will enable us, once we understand early universe conditions to add more substance to the suggestions by Bonnor , as of 1997 for gravity based propulsion systems [13].  As well as permit de facto engineering work pertinent to power source engineering for this concept to become a space craft technology.

# APPENDIX 1A: FIRST TABLE WITH RESPECT TO PHENOMENOLOGY

| Time $0 \leq t \ll t_P$ | Use of quantum gravity to give thermal input via quantum bounce from prior universe collapse to singularity. Brane theory predicts beginning of graviton production. | Axion wall dominant feature of pre inflation conditions, due to Jeans inequality with enhanced gravitational field/ gravitons, **Quintessence scalar equation of motion valid for short time interval** | Wheeler formula for relic graviton production beginning to produce gravitons due to sharp rise in temperatures. |
|---|---|---|---|
| Time $0 \leq t < t_P$ | End of thermal input from quantum gravity due to prior universe quantum bounce. Brane theory predicts massive relic graviton production | Axion wall is in process of disappearing due to mark rise in temperatures. **Quintessence scalar equation of motion valid for short time interval** | Wheeler formula for relic graviton production produces massive spike gravitons due to sharp rise in temperatures within a preferred parameter range. |
| Time $0 < t \approx t_P$ | Relic graviton production largely tapering off, due to thermal input rising above a preferred level, via brane theory calculations. Beginning of regime where four dimensional cosmological constant is approaching values associated with Guth style inflation. See five and four dimensional cosmological constant as given by Park | Axion wall disappears, and beginning of Guth style inflation. **Quintessence scalar equation of motion valid for short time interval. Beginning of regime for which** $\left(\frac{\Lambda}{|\Lambda_5|} - 1\right) \approx O\left(\frac{1}{n}\right)$, i.e. **5 dim** $\to$ **4 dim** | Wheeler formula for relic graviton production leading to few relic gravitons being produced. |
| Time $t > t_P$ | No relic graviton production. Brane theory use scalar potential as given by Sago, et. al. | Approaching regime for which super nova survey called Essence applies. **4 dim only** | Wheeler formula for relic graviton production gives **no** gravitons. |

# APPENDIX 1 B: 2$^{ND}$ TABLE, WHAT CAN BE SAID ABOUT COSMOLGICAL $\Lambda$ IN 5 & 4 DIM

| Time $0 \leq t \ll t_P$ | Time $0 \leq t < t_P$ | Time $0 < t \approx t_P$ | Time $t > t_P \to$ today |
|---|---|---|---|
| $\|\Lambda_5\|$ undefined, $T \approx \varepsilon^+$ | $\|\Lambda_5\| \approx \varepsilon^+$, $T \approx 10^{12} K$ | $\|\Lambda_5\| \approx \Lambda_{4-dim}$, $T$ smaller | $\|\Lambda_5\| \approx$ huge, $T \approx 3.2K$ |
| $\Lambda_{4-dim}$ undefined, $T \approx \varepsilon^+$ | $\Lambda_{4-dim}$ huge, $T \approx 10^{12} K$ | $\Lambda_{4-dim} \approx \|\Lambda_5\|$, $T$ smaller | $\Lambda_{4-dim} \approx 10^{-120} \Lambda_{QCD}$, $T \approx 3.2K$ |

# APPENDIX II: HOW TO DEFINE A TEMPERATURE DEPENDENT FOUR AND FIVE DIMENSIONAL COSMOLOGICAL CONSTANT PARAMETER

We define, via Park's article, [50]

$$k^* = \left(\frac{1}{\text{'AdS curvature}}\right) \tag{1}$$

Park et al note that if we have a 'horizon' temperature term

$$U_T \propto (external\ temperature) \tag{2}$$

We can define a quantity

$$\varepsilon^* = \frac{U_T^4}{k^*} \tag{3}$$

Then there exists a relationship between a four-dimensional version of the $\Lambda_{eff}$, which may be defined by noting

$$\Lambda_{5-dim} \equiv -3 \cdot \Lambda_{4-dim} \cdot \left(\frac{U_T}{k^{*3}}\right)^{-1} \propto -3 \cdot \Lambda_{4-dim} \cdot \left(\frac{external\ temperature}{k^{*3}}\right)^{-1} \tag{4}$$

So

$$\Lambda_{5-dim} \xrightarrow[external\ temperature \to small]{} \text{Large value} \tag{5}$$

And set

$$|\Lambda_{5-dim}| = \Lambda_{eff} \tag{6}$$

# APPENDIX III: THE FOUR VARIETIES OF SHORT LIVED QUINTESSENCE, UP TO A PLANCK'S TIME INTERVAL

**CASE I:**

**Temperature T very small, a.k.a. Carroll and Chen's suppositions (also see Penrose's version of the Jeans inequality) and time less than $t_P$. This is the slow roll case, which is also true when we get to time $\gg t_P$**

$$\ddot{\phi} \cdot \left[1 + \frac{\tilde{c}}{M^2} \cdot \frac{T^2}{6} \cdot g_b\right] + 3 \cdot H \cdot \dot{\phi} \cdot \left[1 + \frac{\tilde{c}}{M^2} \cdot \frac{T^2}{6} \cdot g_b\right] + \left(\frac{\partial V_{axion-contri}}{\partial \phi}\right)$$
$$\xrightarrow[T \to 0^+]{} 3 \cdot H \cdot \dot{\phi} \cdot + \left(\frac{\partial V_{axion-contri}}{\partial \phi}\right) \cong 0 \tag{1}$$

**CASE II:**

**Temperature T very large and time in the neighborhood of $t_P$. This is NOT the slow roll case, and has $H \propto 1/t_P$. Note, which is important that the constant $c$ is not specified to be a small quantity**

$$\ddot{\phi} \cdot \left[1 + \frac{\tilde{c}}{M^2} \cdot \frac{T^2}{6} \cdot g_b\right] + 3 \cdot H \cdot \dot{\phi} \cdot \left[1 + \frac{\tilde{c}}{M^2} \cdot \frac{T^2}{6} \cdot g_b\right] + \left(\frac{\partial V_{axion-contri}}{\partial \phi}\right)$$
$$\xrightarrow[T \to 10^{12} \, Kelvin]{} \ddot{\phi} + 3 \cdot H \cdot \dot{\phi} \cdot + \left(\frac{\tilde{c}}{M^2} \cdot \frac{T^2}{6} \cdot g_b\right)^{-1} \cdot \left(\frac{\partial V_{axion-contri}}{\partial \phi} = m^2 \cdot (\phi - \phi_C)\right) \cong 0 \tag{2}$$

We then get a general, and a particular solution with $\phi_{general} \propto \exp(p \cdot t)$, $\phi_{particular} \equiv \phi_C$, $\phi_{Total} = \phi_{general} + \phi_{particular}$,

$$p^2 + 3 \cdot H \cdot p \cdot + \left(\frac{\tilde{c}}{M^2} \cdot \frac{T^2}{6} \cdot g_b\right)^{-1} \cdot (m^2) \cong 0$$

$$\to p \cong \left[-\frac{3H}{2} \cdot \left[2 - 4 \cdot \frac{m^2 \cdot M^2}{T^2 \cdot c \cdot g_b H}\right], -\left(6 \cdot \frac{m^2 \cdot M^2}{T^2 \cdot c \cdot g_b}\right) \approx -\varepsilon^+\right] \equiv [p_1, p_2] \tag{3}$$

$$\Rightarrow \phi_{general} \cong c_1 \cdot \exp(-|p_1| \cdot t) + c_2 \cdot \exp(-(|p_2| \approx \varepsilon^+) \cdot t)$$

$$\phi_{Total} = \phi_{general} + \phi_{particular} \cong \phi_C + \varepsilon_1 \cdot \phi_{initial \ value} + H.O.T. \ , \text{ where } \varepsilon_1 < 1 \tag{4}$$

**CASE III:**

**Temperature T very large and time in the neighborhood of $t_P$. This is NOT the slow roll case, and has $H \propto 1/t_P$. Note, which is important that the constant $c$ IS specified to be a small quantity. We get much the same analysis as before except the higher order terms (H.O.T.) do not factor in**

$$\phi_{Total} = \phi_{general} + \phi_{particular} \cong \phi_C + \varepsilon_1 \cdot \phi_{initial \ value} \ , \text{ where } \varepsilon_1 < 1 \tag{5}$$

**Case IV: Temperature T not necessarily large but on the way of becoming large valued, so the axion mass is not negligible, YET, and time in the neighborhood of $t_P$. This is NOT the slow roll case, and has $H > H_{t=t_P} \propto 1/t_P$. Begin with making the following approximation to the Axion dominated effective potential**

$$V_{axion-contri} \equiv f[m_{axion}(T)] \cdot (1 - \cos(\phi)) + \frac{m^2}{2} \cdot (\phi - \phi_C)^2 \Rightarrow$$

$$\left(\frac{\partial V_{axion-contri}}{\partial \phi}\right) \xrightarrow{Temperature\ getting\ l\arg er} \quad (6)$$

$$f[m_{axion}(T)] \cdot \frac{\phi^5}{125} - f[m_{axion}(T)] \cdot \frac{\phi^3}{6} + \left[(m^2 + f[m_{axion}(T)]) \cdot \phi - m^2 \phi\right]$$

Then we obtain

$$\ddot{\phi} + 3 \cdot H \cdot \dot{\phi} + \left(\frac{\tilde{c}}{M^2} \cdot \frac{T^2}{6} \cdot g_b\right)^{-1} \cdot \left(\frac{\partial V_{axion-contri}}{\partial \phi}\right) \cong 0 \quad (7)$$

This will lead to as the temperature rises we get that the general solution has definite character as follows

$$p^2 + 3 \cdot H \cdot p + \left(\frac{\tilde{c}}{M^2} \cdot \frac{T^2}{6} \cdot g_b\right)^{-1} \cdot (m^2) \cong 0$$

$$\rightarrow p \cong \left[-\frac{3H}{2} \cdot \left[1 \pm \sqrt{1 - \frac{6 \cdot M^2}{3 \cdot T^2 \cdot c \cdot g_b H} \cdot (m^2 + f[m_{axion}(T)])}\right]\right] \equiv [p_1, p_2]$$

$$\Rightarrow \phi_{general} \cong c_1 \cdot \exp(p_1 \cdot t) + c_2 \cdot \exp(p_2 \cdot t) \quad (8)$$

$$\propto [\phi(real) + i \cdot \phi(imaginary)] \quad iff\ [m_{axion}(T)]\ l\arg e$$

$$\propto [\phi(real)] \quad iff\ [m_{axion}(T)]\ small$$

The upshot is, that for large, but shrinking axion mass contributions we have a cyclical oscillatory system, which breaks down and becomes a real field if the axion mass disappears. First of all, though, we have to understand how the conditions presented by S. Carroll, and J. Chen came about via brane theory. Our point is that we need brane theory to establish the initial starting point of reference for low temperature conditions at the onset of inflation.